\pdfoutput=1 
\documentclass{svproc}
\usepackage{url}
\usepackage{float}
\usepackage{amsmath}
\usepackage{amssymb}
\usepackage{graphicx}

\begin{document}
\mainmatter              % start of a contribution

\title{Dense pedestrian crowds versus granular packings: An analogy of sorts}

\titlerunning{Dense crowds vs. granular packings}  % abbreviated title (for running head)
\author{Alexandre NICOLAS}
\authorrunning{Alexandre NICOLAS} % abbreviated author list (for running head)
\tocauthor{Alexandre NICOLAS}
\institute{Institut Lumi\`ere Mati\`ere, CNRS and Universit\'e Claude Bernard Lyon 1, F-69622 Villeurbanne, France\\ 
     ORCiD: 0000-0002-8953-3924 \\
\email{alexandre.nicolas@polytechnique.edu}}

\maketitle              % typeset the title of the contribution

\begin{abstract}
Analogies between the dynamics of pedestrian crowds and granular media have long been hinted at.
They seem all the more promising as the crowd is (very) dense, in which case the mechanical constraints prohibiting overlaps
might prevail over the decisional component of pedestrian dynamics. These analogies and their origins are probed in two distinct 
settings, (i) a flow through a narrow bottleneck and (ii) crossing of a static assembly by an intruder. Several quantitative 
similarities have been reported for the former setting and are discussed here, while setting (ii) reveals discrepancies in the response pattern, which are
ascribed to the pedestrians' ability to perceive, anticipate and self-propel.

\end{abstract}

Besides its obvious practical relevance, the study of pedestrian dynamics
owes much of its interest to its singular positioning at the crossroads
between social sciences and mechanics. This is reflected by a dichotomy of approaches.

On the one hand, statistical models inspired from economics, such as
discrete-choice models\footnote{These models depend on an abstract utility function, which quantifies the attractiveness of
a given destination, direction or trajectory.}, are often used when it comes to predicting distant destinations or routes chosen by pedestrians
\cite{bierlaire2009pedestrians}. (Nevertheless, it should be mentioned that they have also been applied to stepping dynamics \cite{antonini2006discrete}).

On the other hand, mechanical approaches
are generally favoured when the dynamics are investigated in finer detail, at a more
'microscopic' scale, in particular when strong interactions or hindrances
are expected between pedestrians because of the crowd's density. These approaches are formally based on Newton's equation
 of motion (or variants thereof), supplemented with a
term describing the (externally inferred) desired direction.

How the pedestrians' \emph{decisions} and \emph{choices} of motion in view of their surroudings should enter this mechanical
framework remains however unclear, from a fundamental viewpoint. 
\emph{In practice}, the two components (the 'cognitive' one and the
purely mechanical one) are often combined in an \emph{ad hoc} way.
In the famous social-force model \cite{helbing1995social}, at
the heart of various commercial software products, these components
are simply put on an equal footing, in that the choices of motion
(influenced by the other pedestrians and the built environment) are
converted into 'social forces' and summed up with the mechanical
forces in Newton's equation of motion. One is left with a mechanical
problem, without further justification.

Mechanical descriptions are quite appealing, especially at high density. Indeed,  when the
crowd is very dense, one may even think
that excluded-volume effects (precluding overlaps) restrict the possibilities
of motion so much that they stifle the effect of decision-making,
thus the singularities of pedestrians as compared to, say, grains
of matter. To translate this idea into a reasoning on configuration space, most configurations
are admissible for a sparse crowd, because randomly positioned pedestrians
seldom overlap; in contrast, at very high density,
volume exclusion reduces the available volume of configuration space; the prevalence of
areas that are forbidden because of overlaps might even restrict the possible directions of evolution
of the assembly in configuration space to such an extent that the former virtually govern its evolution (as if in a maze).

This begs the following question: Is a purely mechanical description satisfactory
at very high density? Does the crowd then behave like a granular
medium? We shall address this question in two distinct situations, a bottleneck setup in Section II and an intrusion experiment
in Section III, after
setting the theoretical framework in Section I.

This manuscript aims to provide a pedagogic discussion. Thus, most
quantitative features are skimmed off; references to already published
or forthcoming manuscripts will be provided for readers interested
in quantitative aspects. 

\section{Theoretical framework}

\begin{figure}[H]
\begin{centering}
\includegraphics[width=1\textwidth]{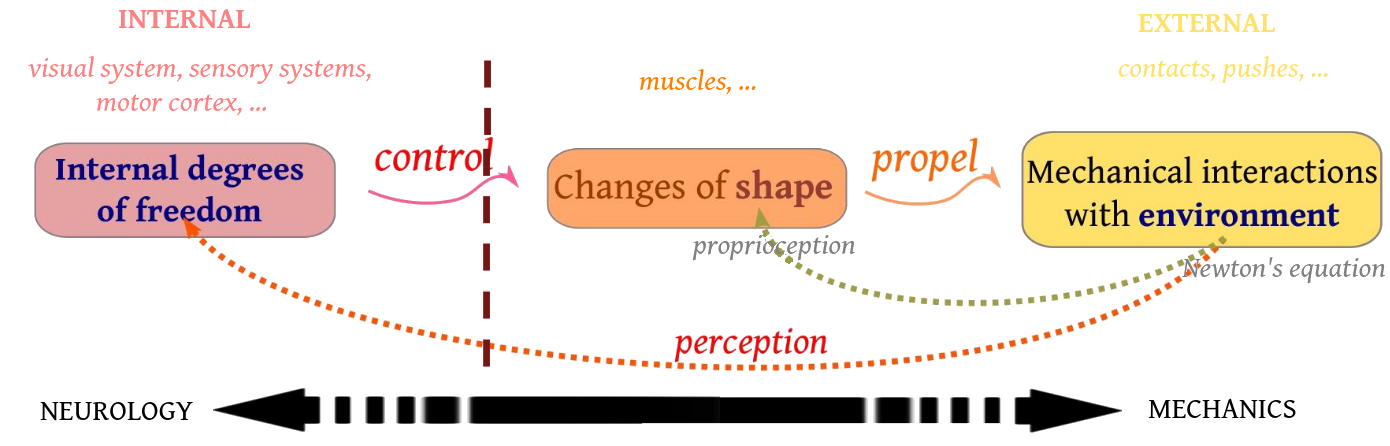}
\par\end{centering}
\caption{\label{fig:sketch}Schematic representation of the interplay between
external mechanical forces, the self-propelling force, and non-mechanical
cues from the environment.}
\end{figure}

We start with a brief introduction to the theoretical framework in
which the motion of pedestrians is to be handled, within classical
mechanics. As a mechanical body, a pedestrian \emph{i }obeys Newton's
equation of motion, 
\[
m\boldsymbol{\ddot{r}}_{i}=\boldsymbol{F}_{\mathrm{ground}\rightarrow i}+\boldsymbol{F}_{\mathrm{wall}\rightarrow i}^{\mathrm{(mech)}}+\sum_{j}\boldsymbol{F}_{j\rightarrow i}^{\mathrm{(mech)}}.
\]
Here, the \emph{mechanical} forces possibly exerted by walls, $\boldsymbol{F}_{\mathrm{wall}\rightarrow i}^{\mathrm{(mech)}}$,
and other pedestrians \emph{j}, $\boldsymbol{F}_{j\rightarrow i}^{\mathrm{(mech)}}$,
in case of contact and pushes, have been taken into account, even
though they seldom arise in practice. The ground force, $\boldsymbol{F}_{\mathrm{ground}\rightarrow i}$,
is composed of (i) a static (vertical) reaction force that would subsist
with no deformation of the pedestrian's body (\emph{passive case)},
and (ii) a force $\boldsymbol{F^{(p)}}_{i}$ that results from changes
in the body shape (created by flexing leg muscles) and that propels
the pedestrian. In the horizontal plane, one can thus write
\[
m\boldsymbol{\ddot{r}}_{i}=\boldsymbol{F^{(p)}}_{i}+\boldsymbol{F}_{\mathrm{wall}\rightarrow i}^{\mathrm{(mech)}}+\sum_{j}\boldsymbol{F}_{j\rightarrow i}^{\mathrm{(mech)}}.
\]
Note that this self-propelling force $\boldsymbol{F^{(p)}}_{i}$ does
not require any process akin to decision-making: Active particles
such as vibration-driven 'hexbugs' \cite{barois2019characterization,patterson2017clogging} are also propelled by $\boldsymbol{F^{(p)}}_{i}$.
Entities that make decisions about their motion on the basis of their
perceptions \cite{moussaid2011simple,Karamouzas2014universal} are
nonetheless singular in that their (active) propelling force will
be a function of \emph{complex dependences on environmental cues, the history
of past body shapes}, etc., generically denoted by '(...)'; see Fig.~\ref{fig:sketch}. Noting that the reaction
to external cues cannot be instantaenous, it needs a finite time $\tau_{\psi}>0$
to be processed (for humans, voluntary reactions take at least 0.1
s; for instance, the behavioural response to a complex visual stimulus
occurs after typically $\tau_{\psi}=0.4\,\mathrm{s}$ \cite{thorpe1996speed}
), one can write
\begin{equation}
\begin{cases}
m\boldsymbol{\ddot{r}}_{i} & =\boldsymbol{F^{(p)}}_{i}+\boldsymbol{F}_{\mathrm{wall}\rightarrow i}^{\mathrm{(mech)}}+\sum_{j}\boldsymbol{F}_{j\rightarrow i}^{\mathrm{(mech)}}\\
\tau_{\psi}\boldsymbol{\dot{F}^{(p)}}_{i} & =f(\ldots)
\end{cases}\label{eq1}
\end{equation}

Of course, the foregoing derivation of the singularity of perceptual entities is 
overly simplified, because $\boldsymbol{F^{(p)}}_{i}$ may in fact
also vary with the configuration of the crowd for non-perceptual entities,
for instance if a magnetic self-propelled particle has its polarisation axis rotated via magnetic interactions with its neighbours. Nonetheless, it is
interesting to note that, even in the case of such an elementary dependence of $\boldsymbol{F^{(p)}}_{i}$
on the environment, the collective behaviour of self-propelled particles may differ from that of their
passive driven counterparts, notably in their self-organisation abilities \cite{bain2017critical}.
A more accurate view on the topic is reserved for a forthcoming paper (Nicolas,
\emph{in preparation}).

In the following, we wonder whether the complex dependences of $f(\ldots)$ in Eq.~\ref{eq1}
could not be overlooked at high densities, due to the prevalence of
the mechanical constraints embodied by the $\boldsymbol{F}^{\mathrm{(mech)}}$'s,
which would thus liken the crowd's response to that of a granular
medium.

\section{Bottleneck flows}

The foregoing question is first addressed in the context of competitive
flows through a (narrow) bottleneck, insofar as this setting seems
to severely constrain the possibilities of motion.

\subsection{Common features}

In the last few years, several similarities (which had long been hinted
at in a tentative way) have been quantitatively demonstrated in experiments
with granular materials flowing out of a vibrated hopper, on the one
hand, and pedestrian crowds asked to rush through a narrow doorway,
on the other hand \cite{Zuriguel2014clogging}. These similarities
include:
\begin{itemize}
\item exponential bursts of egresses in close succession (see \cite{nicolas2018origin}
for a discussion about the universality of this feature)
\item heavy tails in the distribution of time gaps $\tau$ between egresses
in competitive escapes, which are at least reasonably well described
by power laws: $p(\tau)\sim\tau^{_{-\alpha}}$ \cite{Pastor2015experimental}.
Note that this expression entails a divergence of the mean time gap
$\left\langle \tau\right\rangle $ when $\alpha<2$, hence a dwindling
flow rate as the system size tends to infinity.
\item the possibility of a faster-is-slower effect in the evacuation, whereby
tuning up the entities' drive to escape (\emph{by tilting the hopper
or prescribing a more competitive behaviour to the pedestrians}) may
actually delay the evacuation due to longer clogs \cite{Pastor2015experimental}.
It should however be noted that, while the possibility of such an
effect has been experimentally demonstrated with crowds, it requires
high competitiveness and perhaps also aggressive pushes; for less
competitive evacuations, a more trivial faster-is-faster effect is
observed \cite{nicolas2017pedestrian,haghani2019push}
\end{itemize}
Given that the flow rate $J$ is given by $J=1/\left\langle \tau\right\rangle$,
the distribution $p(\tau)$ of time gaps $\tau$ is a central feature,
which we now study in granular flows and then in their pedestrian
counterparts.

\subsection{Granular hopper flows}

Long time gaps are caused by the formation of clogs, due to arches
blocking the aperture of the vibrated hopper in two dimensions. The
ultimate shattering of an existing arch, if it occurs, owes mostly
to the vibration-induced destabilisation of the weakest link in the
arch, i.e., of the grain that forms the widest angle with its neighbours
in the arch \cite{lozano2012breaking}. Thus, we were led to study the evolution of the position
of this grain under vibrations. Using non-trivial, but reasonable
assumptions, we showed that this evolution can be likened to that of a Brownian particle
in an energy trap, with an inverse temperature $\beta$ associated
with the acceleration $\Gamma$ of the vibrations, $\beta\approx\frac{\gamma}{\Gamma^{2}}$,
where $\gamma$ is a drag coefficient \cite{nicolas2018trap}. It
follows from this formal mapping that the lifetime of an arch
is given by the Kramers' escape time
\begin{equation}
\tau\propto\exp\left(\beta\,E_{b}\right),\label{eq:Kramers}
\end{equation}
where $E_{b}$ is the trap depth (or energy barrier). The values of
$E_{b}$ were inferred from ramp experiments, in which the intensity
of vibrations was gradually increased until an existing arch broke;
they were shown to follow a Weibull distribution
\[
p(y)=e^{-y}\,\mathrm{with}\,y\propto\sqrt{E_{b}}
\]

Arch stabilities, quantified by $E_{b}$, are thus fairly narrowly
distributed. But the bottomline is that this fairly narrow disorder
is amplified by the exponential dependence on $E_{b}$ in Kramers'
escape time, eq.~\ref{eq:Kramers}. Were energy barriers exponentially
distributed (as they are in the Soft Glassy Rheology model \cite{Sollich1997}),
this would lead to a power-law distribution of escape times $\tau$.
Here, the disorder in the arch stabilities is slightly larger, therefore
the distribution of arch lifetimes decays even more slowly than a
power law as $\tau\to\infty$. The results of the foregoing minimal
model attain semi-quantitative agreement with experimental data \cite{nicolas2018trap}.
Importantly, they lead to the prediction that the mean time gap $\left\langle \tau\right\rangle $
may diverge irrespective of the vibration intensity (in the limit
of moderate shaking), which casts doubt on the ability of gentle vibrations
to restore the hopper flow persistently.

\subsection{Pedestrian flows through a bottleneck}

Turning to pedestrians, temporary clogs are also observed in competitive
pedestrian flows through a bottleneck, due to pedestrian 'arches'
blocking the doorway. Our
initial observation was that cellular automata (CA) provide a computationally
very efficient way to simulate them, e.g., to test evacuation scenarios,
but that their predictions are seldom validated against experimental
measurements, in particular detailed statistics about the time series
of egresses \cite{Pastor2015experimental}. As a matter of fact, it seemed to us that
none of the CA that we surveyed in the literature was able to describe
the seemingly heavy-tailed distribution of time gaps between egresses reported for
highly competitive evacuations \cite{nicolas2016statistical}. This
feature turned out to be difficult to replicate. Indeed, the regular
grid typically used in CA precludes geometric disorder in the pedestrian
arches, whereas such disorder was central to the replication of heavy
tails in granular flows (see the previous section).

It goes without saying that the specific mechanics governing the clogging
dynamics differ between pedestrians and grains. However, the broad
insight gained from the granular case helped us devise a CA that was
finally able to reproduce experimental results on controlled evacuations
semi-quantitatively. To this end, after realising (i) the importance of
having some intrinsic disorder in the model (and not just random fluctuations)
and (ii) the constraints affecting the geometry in a CA, we decided to introduce
disorder in the agents' behaviours, more precisely in the impatience
displayed by agents close to the exit (the so called 'friction'),
which controls the number of conflicting endeavours to escape simultaneously.
This enabled us to capture the desired experimental features \cite{nicolas2016statistical}.

All in all, despite the discrepancies in the physical origins of their
motion, (competitive) pedestrians and grains flowing through a narrow
bottleneck exhibit striking similarities, which hints that basic mechanisms
control the dynamics in this setup, in particular the joint presence
of a driving 'force' towards the escape and paramount volume exclusion
effects (included in $\boldsymbol{F}_{j\rightarrow i}^{\mathrm{(mech)}}$
in Eq.~\ref{eq1}) \footnote{Note that these similarities may perhaps not extend to the effect
of placing an obstacle in front of the door \cite{garcimartin2018redefining}. }.

\section{Crossing by an intruder}

To test if the similarities observed in bottleneck flows extend to
other setups, we now move on to crossing tests of a static (pedestrian
or granular) two-dimensional assembly by an intruder.

\begin{figure}[H]
\begin{centering}
\includegraphics[width=1\textwidth]{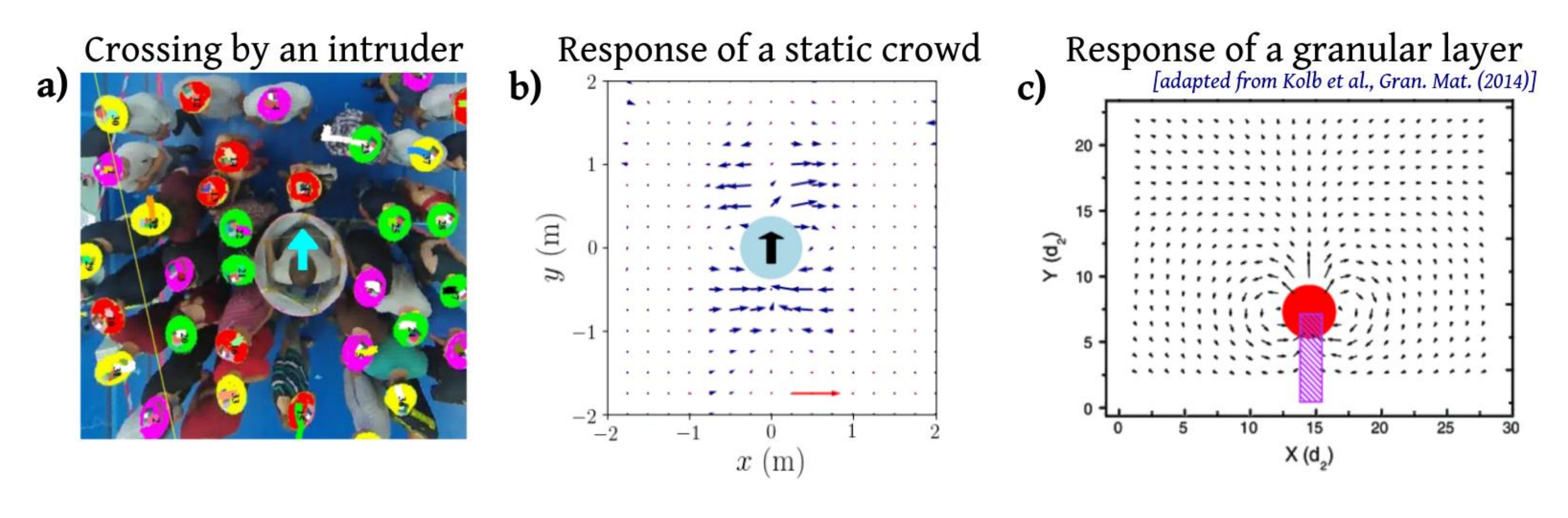}
\par\end{centering}
\caption{\label{fig:response}Response to the crossing of an intruder in a static
crowd (a-b) and a granular layer (c). Picture (a) is a snapshot taken
during a controlled experiment, while panels (b) and (c) are average
flow fields in the co-moving frame of the intruder. The red arrow
in (b) represents a velocity of 30 cm/s. Adapted from Ref. \cite{nicolas2019mechanical,kolb2014flow}.}
\end{figure}

\subsection{Granular mono-layer}

This kind of test is fairly classical for granular media. In particular,
Cixous, Kolb and colleagues studied the passage of a circular intruder
in a mono-layer of granular disks below jamming \cite{kolb2014flow}
(also see the related works by S\'eguin et al. \cite{Seguin2013}).
They distinguished different regimes, but in the regime most relevant
for the upcoming comparison they observed an average flow field of
the form displayed in Fig.~\ref{fig:response}c. It features radial
velocities on the fore-side of the intruder, with recirculation eddies
on the side. Furthermore, the perturbation is short-ranged and decays
exponentially in space (contrary to what happens in a viscous fluid).
It should also be stated that a depleted zone appears in the wake
of the cylinder, with a size that decreases with increasing packing
fraction. 

\subsection{Static pedestrian crowd}

To allow comparison, controlled experiments were performed in Orsay
(France) and Bariloche (Argentina), in which a cylinder of around
70 cm in diameter moved linearly through a group of static participants
(see Fig.~\ref{fig:response}a). The group's density was varied from $1-2\mathrm{\,ped/m^2}$ to $6-7\mathrm{\,ped/m^2}$ and its orientation was also varied; for further experimental details, refer
to \cite{nicolas2019mechanical}. 

In short, when the participants were asked to behave casually, as
if they were on an underground platform, and were facing the incoming intruder,
the crowd's response consisted of strictly transverse moves directed outwards
downstream from the obstacle and inwards in its wake, as shown in
Fig.~\ref{fig:response}b, with only little sensitivity to the crowd's
density. This is in stark contrast with the granular response of Fig.~\ref{fig:response}c.
On the other hand, similarly to the granular case, the perturbation
was short-ranged and left a pedestrian-free zone in the wake of the
obstacle, which was refilled after the passage, via the inwards moves
mentioned just above.

Only when the participants were asked to refrain from anticipating
the intruder's passage \emph{and} had to turn their back to the intruder
(which thus arrived from behind), only then did the crowd's response mirror the
granular one somewhat more closely, with displacements aligned approximately
radially on the fore-side of the intruder (\emph{data not shown here}).

\subsection{Pedestrian specificities}

As in the bottleneck setup, but to a much lesser extent here, some
similarities were noticed in the pedestrian and granular responses, which consist of
a short-ranged perturbation field (possibly due to the discreteness
of the systems) and a depleted zone in the intruder's wake. In contrast,
the observed discrepancies in the displacement patern are ascribed to the following two salient specificities of
pedestrians, as compared to grains, which were also highlighted in the
schematic diagram of Fig.~\ref{fig:sketch}.

Firstly, self-propulsion grants pedestrians the possibility to move
along a direction that is not aligned with the mechanical force applied
by their surroundings (i.e., the intruder and the crowd). Secondly,
the internal 'decision-making' process informed by the perception
of the environment allows pedestrians to anticipate the passage of
the intruder, hence start moving before contact and elect a transverse
move, rather than an axial one (so as to dodge out of the intruder's
way, instead of delaying the collision).

In the formal framework outlined in Section~1, this is tantamount
to saying that an avoidance strategy is encoded in the function $f(\ldots)$
determining the evolution of the self-propelling force in Eq.~\ref{eq1};
this strategy is not reducible to a simple social or mechanical force
inducing radial repulsion. It remains to be checked whether the observed
features (beyond the reach of simple social force models) can be captured
by some of more elaborate lines of modelling, which provide a more
realistic account of anticipating effects, in particular descriptions
based on anticipated times to collision \cite{Karamouzas2014universal}.

\bibliographystyle{spphys}
\bibliography{/home/alexandre/Documents/BiblioCrowds,/home/alexandre/Documents/PhDbib}

\end{document}